# Uniform electric field induced lateral migration of a sedimenting drop

Aditya Bandopadhyay[1], Shubhadeep Mandal[2] and Suman Chakraborty[1,2]†

[1]Advanced Technology Development Center, Indian Institute of Technology Kharagpur, West Bengal- 721302, India

[2]Department of Mechanical Engineering, Indian Institute of Technology Kharagpur, West Bengal- 721302, India

We investigate the motion of a sedimenting spherical drop in the presence of an applied uniform electric field in an otherwise arbitrary direction in the limit of low surface charge convection. We analytically solve the electric potential in and around the leaky dielectric drop, and solve for the Stokesian velocity and pressure fields. We obtain the drop velocity through perturbations in powers of the electric Reynolds number which signifies the importance of the charge relaxation time scale as compared to the convective time scale. We show that in the presence of electric field either in the sedimenting direction or orthogonal to it, there is a change in the drop velocity only in the direction of sedimentation due to an asymmetric charge distribution in the same direction. However, in the presence of an electric field applied in both the directions, and depending on the permittivities and conductivities of the two fluids, we obtain a non-intuitive lateral migration of drop in addition to the buoyancy driven sedimentation. These dynamical features can be effectively used for manipulating drops in a controlled electro-fluidic environment.

______________________________________________________________________

## 1. Introduction

The study of lateral migration of particles (solid particles, drops, bubbles and biological cells) is of immense importance because of its practical uses in microfluidic applications such as sorting and separation of particles, transport of biological cells and so on (Bhagat et al. 2010; Di Carlo et al. 2007; Chen 2014; Geislinger & Franke 2013). Different mechanisms that have shown to cause the lateral migration of particles in the presence of background flows pertain to interface deformability (Haber & Hetsroni 1972; Mortazavi & Tryggvason 2000; Griggs et al. 2007), fluid inertia (Ho & Leal 2006), non-Newtonian rheology (Haber & Hetsroni 1971; Aggarwal & Sarkar 2007; Mukherjee & Sarkar 2013), and Marangoni effects (Hanna & Vlahovska 2010; Pak et al. 2014). Recently, the use of an external electric field to manipulate particles has been employed due to the ease with which the electric field can be applied and modulated (McGrath et al. 2014; Velev & Bhatt 2006; Velev et al. 2003; Zeng & Korsmeyer 2004).

Electrohydrodynamics of drops and bubbles has been studied extensively in the presence of electric fields. Starting from the pioneering work of Taylor (1966), several

†Email address for correspondence: suman@mech.iitkgp.ernet.in

...authors have addressed the problem of deformation of neutrally buoyant leaky dielectric drops in uniform and non-uniform electric fields (Vizika & Saville 2006; Torza et al. 1971; Supeene et al. 2008; Lanauze et al. 2013; Zhang et al. 2013; Deshmukh & Thaokar 2013; Thaokar 2012; Lac & Homsy 2007; Feng & Scott 2006; Ward & Homsy 2006; Ward & Homsy 2003; Ward & Homsy 2001). In these works it was implicitly considered that the flow field does not affect the charge distribution at the surface and this assumption decouples the electric field from the flow field. As a non-trivial extension of the previously mentioned works, Feng (1999) demonstrated that the deformation is significantly altered by the surface charge convection. From this discussion one would expect that a background flow field may also significantly affect the drop dynamics. However, there is very little study (Xu & Homsy 2006; Vlahovska 2011; Xu 2007; Mahlmann & Papageorgiou 2009; He et al. 2013) in the present literature which considers the electrohydrodynamic motion of a single drop or bubble in the presence of an imposed background flow, which is very common in modern microfluidic devices. Xu & Homsy (2006) have shown the influence of surface charge convection on the velocity of a sedimenting leaky dielectric drop in the presence of uniform axial electric field for nearly spherical drop and small charge relaxation time scale. Later, Vlahovska (2011) studied the drop dynamics in the presence of imposed shear flow and uniform electric field, and showed that the surface charge convection alters the rotation rate of a highly viscous drop. The author also showed that the surface charge convection affects the shear viscosity and normal stress of a dilute emulsion containing drops.

In the present work, our focus is to bring out the effect of surface charge convection on the drop motion in the presence of an *arbitrarily oriented* uniform electric field. We *analytically* derive the motion of a sedimenting spherical drop and the associated electric and flow fields in the presence of the electric field. We first show that the sedimentation velocity is affected when the applied electric field acts in the direction of sedimentation (longitudinal) or acts in the orthogonal direction of sedimentation (transverse). Quite remarkably, our results reveal a non-intuitive transverse migration of drop due to a simultaneous application of electric field in both longitudinal and transverse directions. The direction of migration depends on the ratios of the permittivities and conductivities of the two fluids. In sharp contrast to the case of a uniform axial electric field and contrary to common intuition, our results reveal the droplet dynamics due to the central role of the coupling between the direction of the applied electric field and the impending charge convection brought about by the drop sedimentation. Thus, our analysis extends the axisymmetric considerations by Xu and Homsy (2006) of a sedimenting drop in the presence of a longitudinal electric field to a significantly more generalized paradigm, and aims at providing a unified framework for motion of drops in arbitrary uniform electric field.

**2. Problem statement**

We consider a Newtonian leaky dielectric drop of radius *a* suspended in another Newtonian leaky dielectric fluid sedimenting under the influence of gravity in the positive *z* direction with an externally applied electric field $\left(\mathbf{E}_\infty\right)$ in an arbitrary direction (please refer to figure 1). In the present work, we assume that the Reynolds number is sufficiently small so as to completely ignore the inertia in the governing equations for fluid flow. It is also assumed that the interfacial surface tension is sufficiently large to prevent any significant deformation from sphericity. The center of the spherical coordinate system $(r, \theta, \phi)$ is chosen at the center of the drop which moves with a, yet unknown, velocity $\mathbf{U}_d$.

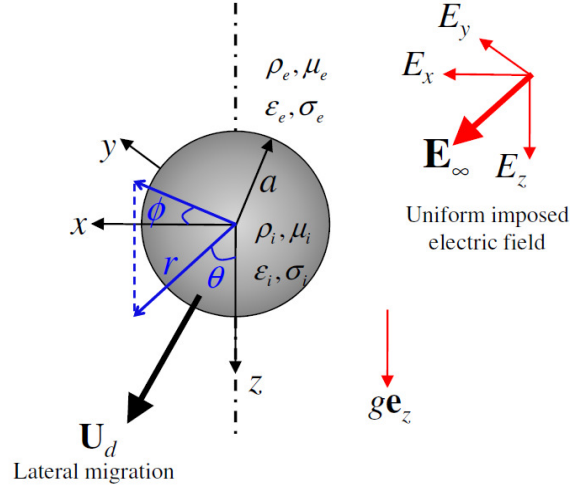

FIGURE 1. (Colour online) Schematic of a drop sedimenting in the presence of an imposed electric field $\mathbf{E}_\infty \left(= E_x \mathbf{e}_x + E_y \mathbf{e}_y + E_z \mathbf{e}_z \right)$ in an arbitrary direction. The density, viscosity, permittivity and conductivity of the inner drop are denoted by $\rho_i, \mu_i, \varepsilon_i$ and $\sigma_i$ respectively; while the same symbols with subscript $e$ denote the properties of the external fluid.

We non-dimensionalize the length by the drop radius $a$. In the absence of any applied electric field, the drop sediments in the $z$ direction with a sedimenting velocity given by the Hadamard-Rybczynski velocity (Leal 2007) $U_{HR} = \frac{2}{9} \frac{ga^2(\rho_i - \rho_e)}{\mu_e} \frac{(\lambda+1)}{(\lambda+\frac{2}{3})}$, where $\lambda$ denotes the drop to suspending medium viscosity ratio. Following the arguments by Xu and Homsy (2006), we choose the characteristic velocity as $U_{HR}$ and correspondingly, the characteristic hydrodynamic stress is obtained as $\tau_{ref}^H = \mu_e U_{HR}/a$. On similar lines, the electrical stress is non-dimensionalized by $\tau_{ref}^E = \varepsilon_e E_{ref}^2$ with $E_{ref}$ being the characteristic magnitude of the applied electric field. In the present work, all the other properties are scaled by the properties of the outside fluid; $\lambda (= \mu_i/\mu_e)$ denotes the viscosity ratio of the drop to medium, $R (= \sigma_i/\sigma_e)$ denotes the conductivity ratio of the drop to medium and $S (= \varepsilon_i/\varepsilon_e)$ denotes the permittivity ratio of the drop to medium. Under these considerations, we obtain the following important non-dimensional parameters (Vlahovska 2011; Xu & Homsy 2006): $M = \frac{a\varepsilon_e E_{ref}^2}{\mu_e U_{HR}}$ and $Re_E = \frac{\varepsilon_e U_{HR}}{a\sigma_e}$. Here, $M$ is the Mason parameter which signifies the ratio of the electrical stress to the viscous stress; and $Re_E$ is the electric Reynolds number which is the ratio of the charge relaxation timescale to the convective timescale.

### 2.1. *Governing equations and boundary conditions*

Throughout our analysis, we present the governing equations and the respective boundary conditions in a non-dimensional form unless specified otherwise. Under the consideration of the leaky dielectric model (Melcher & Taylor 1969; Taylor 1966), bulk fluids are charge free and the electric potentials, $\psi_{i,e}$, satisfy the Laplace equation of the form

$$\nabla^2 \psi_i = 0, \qquad (2.1)$$
$$\nabla^2 \psi_e = 0.$$

The electric potential $(\psi_{i,e})$ can be obtained by solving equation (2.1) by using the following boundary conditions (Xu & Homsy 2006; Saville 1997):

(e1) specified unperturbed electric field at infinity; at $r \to \infty$, $\psi_e \to \psi_\infty$, where $\mathbf{E}_\infty = -\nabla \psi_\infty$,

(e2) bounded electric field inside the drop,

(e3) continuity of electric potential at the drop interface; at $r = 1$, $\psi_i = \psi_e$,

(e4) charge conservation at the drop interface; at $r = 1$, $\mathbf{n} \cdot (R \nabla \psi_i - \nabla \psi_e) = -Re_E \nabla_s \cdot (q \mathbf{V}_s)$, where $q = \mathbf{n} \cdot (S \nabla \psi_i - \nabla \psi_e)$ is the surface charge distribution, $\mathbf{n} = \mathbf{e}_r$ represents the outward unit normal vector on the spherical drop interface and $\mathbf{V}_s$ is the velocity at the drop interface. Here $\nabla_s = \nabla - \mathbf{n}(\mathbf{n} \cdot \nabla)$ is the surface divergence operator.

The flow field satisfies the Stokes equation and the condition of incompressibility of the form (Happel & Brenner 1981)

$$\nabla p_i = \lambda \nabla^2 \mathbf{u}_i, \quad \nabla \cdot \mathbf{u}_i = 0, \qquad (2.2)$$
$$\nabla p_e = \nabla^2 \mathbf{u}_e, \quad \nabla \cdot \mathbf{u}_e = 0,$$

where $\mathbf{u}_i$ and $p_i$ represent velocity and pressure field inside the drop, respectively, while subscript $e$ has been used to denote the same quantities outside the drop. One thing to note here is that pressure $p_i$ and $p_e$ contain the contribution of the gravitational potential.

In a reference frame attached to the drop centre, the flow field satisfies the following boundary conditions:

(f1) uniform velocity field at infinity; $\mathbf{u}_e(r \to \infty) = \mathbf{u}_\infty = -\mathbf{U}_d$,

(f2) velocity and pressure field inside the drop is bounded,

(f3) velocity fields satisfy the following boundary condition at the interface $(r=1)$: $\mathbf{u}_i = \mathbf{u}_e$ and $\mathbf{u}_i \cdot \mathbf{n} = \mathbf{u}_e \cdot \mathbf{n} = 0$,

(f4) tangential stresses are continuous at the drop interface $(r=1)$; $\mathbf{n} \cdot \boldsymbol{\tau}_i \cdot (\mathbf{I} - \mathbf{nn}) = \mathbf{n} \cdot \boldsymbol{\tau}_e \cdot (\mathbf{I} - \mathbf{nn})$, where $\boldsymbol{\tau}_{i,e}$ represetns the total stress tensor consisting of the hydrodynamic and electrical effects. Note that in (f4), $\mathbf{I} - \mathbf{nn}$ represents the surface projection operator.

## 2.2. *Expansion in perturbation of $Re_E$*

For problems dealing with migration due to deformation, it is typical to perform a regular domain perturbation in orders of capillary number which signifies the relative strength of viscous force as compared to capillary force. However, we would like to focus only on the effect of surface charge convection on the motion of the drop. Towards this, we have assumed that the drop remains spherical, which is in line with some previous analysis (Hanna & Vlahovska 2010; Pak et al. 2014). Here we use electrical Reynolds number $Re_E$ as a perturbation parameter which quantifies the effectiveness of surface convection; at leading order surface convection is absent. Now, we expand the electric potential, velocity, pressure, stress fields and the unknown drop velocity in the following regular perturbation form

$$\begin{aligned}
\psi_{i,e} &= \psi_{i,e}^{(0)} + Re_E \psi_{i,e}^{(1)} + Re_E^2 \psi_{i,e}^{(2)} + \cdots, \\
\mathbf{u}_{i,e} &= \mathbf{u}_{i,e}^{(0)} + Re_E \mathbf{u}_{i,e}^{(1)} + Re_E^2 \mathbf{u}_{i,e}^{(2)} + \cdots, \\
p_{i,e} &= p_{i,e}^{(0)} + Re_E p_{i,e}^{(1)} + Re_E^2 p_{i,e}^{(2)} + \cdots, \\
\boldsymbol{\tau}_{i,e} &= \boldsymbol{\tau}_{i,e}^{(0)} + Re_E \boldsymbol{\tau}_{i,e}^{(1)} + Re_E^2 \boldsymbol{\tau}_{i,e}^{(2)} + \cdots, \\
\mathbf{U}_d &= \mathbf{U}_d^{(0)} + Re_E \mathbf{U}_d^{(1)} + Re_E^2 \mathbf{U}_d^{(2)} + \cdots.
\end{aligned} \quad (2.3)$$

Here $\mathbf{U}_d^{(0)}$ is the drop velocity in the absence of interfacial charge convection and $\mathbf{U}_d^{(1)}$ denotes the first contribution of charge convection on the drop velocity.

Considering the above, the governing equations for the electric field at the leading order are given as:

$$\begin{aligned}
\nabla^2 \psi_i^{(0)} &= 0, \\
\nabla^2 \psi_e^{(0)} &= 0,
\end{aligned} \quad (2.4)$$

subjected to the following boundary conditions

$$\begin{aligned}&\text{at } r \to \infty, \ \psi_e^{(0)} \to \psi_\infty \\ &\psi_i^{(0)} \text{ bounded for } r < 1 \\ &\text{at } r = 1, \ \psi_i^{(0)} = \psi_e^{(0)} \\ &\text{at } r = 1, \ \mathbf{e}_r \cdot \left( R \nabla \psi_i^{(0)} - \nabla \psi_e^{(0)} \right) = 0. \end{aligned} \qquad (2.5)$$

The leading order velocity field and pressure satisfy the following governing equations

$$\begin{aligned} \nabla p_i^{(0)} &= \lambda \nabla^2 \mathbf{u}_i^{(0)}, \quad \nabla \cdot \mathbf{u}_i^{(0)} = 0, \\ \nabla p_e^{(0)} &= \nabla^2 \mathbf{u}_e^{(0)}, \quad \nabla \cdot \mathbf{u}_e^{(0)} = 0, \end{aligned} \qquad (2.6)$$

which are subjected to the following boundary condition

$$\begin{aligned} &\mathbf{u}_e^{(0)}(r \to \infty) = \mathbf{u}_\infty^{(0)} = -\mathbf{U}_d^{(0)} \\ &\mathbf{u}_i^{(0)} \text{ is bounded for } r < 1 \\ &\text{at } r = 1, \ \mathbf{u}_i^{(0)} = \mathbf{u}_e^{(0)} \\ &\text{at } r = 1, \ \mathbf{u}_i^{(0)} \cdot \mathbf{e}_r = \mathbf{u}_e^{(0)} \cdot \mathbf{e}_r = 0 \\ &\text{at } r = 1, \ \mathbf{e}_r \cdot \boldsymbol{\tau}_i^{(0)} \cdot (\mathbf{I} - \mathbf{e}_r \mathbf{e}_r) = \mathbf{e}_r \cdot \boldsymbol{\tau}_e^{(0)} \cdot (\mathbf{I} - \mathbf{e}_r \mathbf{e}_r). \end{aligned} \qquad (2.7)$$

On similar lines, we obtain the following equations for the first order electric field:

$$\begin{aligned} \nabla^2 \psi_i^{(1)} &= 0, \\ \nabla^2 \psi_e^{(1)} &= 0, \end{aligned} \qquad (2.8)$$

which are subjected to the, now altered, first order boundary conditions (Feng 1999; Xu & Homsy 2006)

$$\begin{aligned} &\text{at } r \to \infty, \ \psi_e^{(1)} \to 0 \\ &\psi_i^{(1)} \text{ bounded for } r < 1 \\ &\text{at } r = 1, \ \psi_i^{(1)} = \psi_e^{(1)} \\ &\text{at } r = 1, \ \mathbf{e}_r \cdot \left( R \nabla \psi_i^{(1)} - \nabla \psi_e^{(1)} \right) = -\nabla_s \cdot \left( q^{(0)} \mathbf{V}_s^{(0)} \right) \\ &\text{where } q^{(0)} = \mathbf{e}_r \cdot \left( S \nabla \psi_i^{(0)} - \nabla \psi_e^{(0)} \right). \end{aligned} \qquad (2.9)$$

It is to be noted that in the boundary condition for the normal electric field, the effect of surface convection now appears through the redistribution of the surface charge. The first order velocity field can then be found out using:

$$\begin{aligned} \nabla p_i^{(1)} &= \lambda \nabla^2 \mathbf{u}_i^{(1)}, \quad \nabla \cdot \mathbf{u}_i^{(1)} = 0, \\ \nabla p_e^{(1)} &= \nabla^2 \mathbf{u}_e^{(1)}, \quad \nabla \cdot \mathbf{u}_e^{(1)} = 0, \end{aligned} \qquad (2.10)$$

which are subjected to the boundary conditions

$$\begin{aligned}
&\mathbf{u}_e^{(1)}(r \to \infty) = \mathbf{u}_\infty^{(1)} = -\mathbf{U}_d^{(1)}, \\
&\mathbf{u}_i^{(1)} \text{ is bounded for } r < 1, \\
&\text{at } r = 1, \ \mathbf{u}_i^{(1)} = \mathbf{u}_e^{(1)}, \\
&\text{at } r = 1, \ \mathbf{u}_i^{(1)} \cdot \mathbf{e}_r = \mathbf{u}_e^{(1)} \cdot \mathbf{e}_r = 0, \\
&\text{at } r = 1, \ \mathbf{e}_r \cdot \boldsymbol{\tau}_i^{(1)} \cdot (\mathbf{I} - \mathbf{e}_r \mathbf{e}_r) = \mathbf{e}_r \cdot \boldsymbol{\tau}_e^{(1)} \cdot (\mathbf{I} - \mathbf{e}_r \mathbf{e}_r).
\end{aligned} \quad (2.11)$$

## 3. Analytical solution in the spherical limit

### 3.1. *First iteration - leading order electric field and flow field*

The solution of equation (2.4) which gives the potential distribution inside and outside the drop is of the form:

$$\begin{aligned}
\psi_i^{(0)} &= \sum_{n=0}^{\infty} \left( a_{n,m}^{(0)} \cos m\phi + \hat{a}_{n,m}^{(0)} \sin m\phi \right) r^n P_{n,m}(\cos\theta), \\
\psi_e^{(0)} &= \psi_\infty + \sum_{n=0}^{\infty} \left( b_{n,m}^{(0)} \cos m\phi + \hat{b}_{n,m}^{(0)} \sin m\phi \right) r^{-n-1} P_{n,m}(\cos\theta),
\end{aligned} \quad (3.1)$$

where $\psi_\infty = -r\left(E_x P_{1,1} \cos\phi + E_y P_{1,1} \sin\phi + E_z P_{1,0}\right)$ is the unperturbed imposed electric potential at infinity, $\{E_x, E_y, E_z\}$ are the non-dimensional components of applied electric field and $P_{n,m}(\cos\theta)$ is the associated Legendre polynomial of degree $n$ and order $m$. The unknown coefficients $\left(a_{n,m}^{(0)}, \hat{a}_{n,m}^{(0)}, b_{n,m}^{(0)} \text{ and } \hat{b}_{n,m}^{(0)}\right)$ can be determined by using the boundary conditions (equation(2.5)) and subsequently the leading order electric potential can be obtained as

$$\begin{aligned}
\psi_i^{(0)} &= -\frac{3E_z}{2+R} r P_{1,0} - \left(\frac{3E_x \cos\phi}{2+R} + \frac{3E_y \sin\phi}{2+R}\right) r P_{1,1}, \\
\psi_e^{(0)} &= -r\left(E_x P_{1,1} \cos\phi + E_y P_{1,1} \sin\phi + E_z P_{1,0}\right) + \frac{E_z(R-1)}{(2+R)r^2} P_{1,0} \\
&\quad + \left[\frac{E_x(R-1)\cos\phi}{2+R} + \frac{E_y(R-1)\sin\phi}{2+R}\right] \frac{1}{r^2} P_{1,1}.
\end{aligned} \quad (3.2)$$

The leading order surface charge distribution at the drop interface is obtained as

$$q^{(0)} = \frac{3(R-S)}{R+2}\left[E_z P_{1,0} + \left(E_x \cos\phi + E_y \sin\phi\right) P_{1,1}\right]. \quad (3.3)$$

The general solution of equation (2.6) for the flow field in spherical coordinates can be obtained by using the Lamb solution (Lamb 1975) which expresses the velocity and pressure in terms of solid spherical harmonics. The leading order velocity and pressure field inside the drop can be represented as (Pak & Lauga 2014; Happel & Brenner 1981; Hetsroni & Haber 1970)

$$\mathbf{u}_i^{(0)} = \sum_{n=1}^{\infty} \left[ \nabla \times \left( \mathbf{r} \chi_n^{(0)} \right) + \nabla \Phi_n^{(0)} + \frac{n+3}{2(n+1)(2n+3)\lambda} r^2 \nabla p_n^{(0)} - \frac{n}{(n+1)(2n+3)\lambda} \mathbf{r} p_n^{(0)} \right],$$

$$p_i^{(0)} = \sum_{n=1}^{\infty} p_n^{(0)},$$ 
(3.4)

where $\mathbf{r}$ is the dimensionless position vector of magnitude $r$; $p_n^{(0)}, \Phi_n^{(0)}$ and $\chi_n^{(0)}$ are the growing solid spherical harmonics of the following form

$$p_n^{(0)} = \lambda r^n \sum_{m=0}^{n} \left( A_{n,m}^{(0)} \cos m\phi + \hat{A}_{n,m}^{(0)} \sin m\phi \right) P_{n,m},$$

$$\Phi_n^{(0)} = r^n \sum_{m=0}^{n} \left( B_{n,m}^{(0)} \cos m\phi + \hat{B}_{n,m}^{(0)} \sin m\phi \right) P_{n,m},$$ 
(3.5)

$$\chi_n^{(0)} = r^n \sum_{m=0}^{n} \left( C_{n,m}^{(0)} \cos m\phi + \hat{C}_{n,m}^{(0)} \sin m\phi \right) P_{n,m}.$$

Similarly, the leading order velocity and pressure outside the drop can be represented using Lamb solution as

$$\mathbf{u}_e^{(0)} = \mathbf{u}_\infty^{(0)} + \mathbf{v}_e^{(0)}$$
$$= \mathbf{u}_\infty^{(0)} + \sum_{n=1}^{\infty} \left[ \nabla \times \left( \mathbf{r} \chi_{-n-1}^{(0)} \right) + \nabla \Phi_{-n-1}^{(0)} - \frac{n-2}{2n(2n-1)} r^2 \nabla p_{-n-1}^{(0)} + \frac{n+1}{n(2n-1)} \mathbf{r} p_{-n-1}^{(0)} \right],$$

$$p_e^{(0)} = \sum_{n=1}^{\infty} p_{-n-1}^{(0)},$$ 
(3.6)

where $\mathbf{v}_e^{(0)}$ is the disturbance velocity field outside the drop which vanishes at infinity. Here $p_{-n-1}^{(0)}, \Phi_{-n-1}^{(0)}$ and $\chi_{-n-1}^{(0)}$ are decaying solid spherical harmonics of the following form

$$p_{-n-1}^{(0)} = r^{-n-1} \sum_{m=0}^{n} \left( A_{-n-1,m}^{(0)} \cos m\phi + \hat{A}_{-n-1,m}^{(0)} \sin m\phi \right) P_{n,m},$$

$$\Phi_{-n-1}^{(0)} = r^{-n-1} \sum_{m=0}^{n} \left( B_{-n-1,m}^{(0)} \cos m\phi + \hat{B}_{-n-1,m}^{(0)} \sin m\phi \right) P_{n,m},$$ 
(3.7)

$$\chi_{-n-1}^{(0)} = r^{-n-1} \sum_{m=0}^{n} \left( C_{-n-1,m}^{(0)} \cos m\phi + \hat{C}_{-n-1,m}^{(0)} \sin m\phi \right) P_{n,m}.$$

In a similar manner $\mathbf{u}_\infty^{(0)}$ can also be expressed in terms of spherical harmonics as

$$\mathbf{u}_{\infty}^{(0)} = \sum_{n=-\infty}^{\infty} \left[ \nabla \times \left( \mathbf{r} \chi_n^{\infty(0)} \right) + \nabla \Phi_n^{\infty(0)} + \frac{n+3}{2(n+1)(2n+3)} r^2 \nabla p_n^{\infty(0)} - \frac{n}{(n+1)(2n+3)} \mathbf{r} p_n^{\infty(0)} \right], \tag{3.8}$$

where $p_n^{\infty(0)}, \Phi_n^{\infty(0)}$ and $\chi_n^{\infty(0)}$ are solid spherical harmonics of the following form (Hetsroni & Haber 1970)

$$p_n^{\infty(0)} = \frac{2(2n+3)}{n} r^n \sum_{m=0}^{n} \left( \alpha_{n,m}^{(0)} \cos m\phi + \hat{\alpha}_{n,m}^{(0)} \sin m\phi \right) P_{n,m},$$

$$\Phi_n^{\infty(0)} = \frac{1}{n} r^n \sum_{m=0}^{n} \left( \beta_{n,m}^{(0)} \cos m\phi + \hat{\beta}_{n,m}^{(0)} \sin m\phi \right) P_{n,m}, \tag{3.9}$$

$$\chi_n^{\infty(0)} = \frac{1}{n(n+1)} r^n \sum_{m=0}^{n} \left( \gamma_{n,m}^{(0)} \cos m\phi + \hat{\gamma}_{n,m}^{(0)} \sin m\phi \right) P_{n,m}.$$

The coefficients $\alpha_{n,m}^{(0)}, \hat{\alpha}_{n,m}^{(0)}, \beta_{n,m}^{(0)}, \hat{\beta}_{n,m}^{(0)}, \gamma_{n,m}^{(0)}$ and $\hat{\gamma}_{n,m}^{(0)}$ can be related to the unknown drop velocity at leading order ($\mathbf{U}_d^{(0)} = U_{dx}^{(0)} \mathbf{e}_x + U_{dy}^{(0)} \mathbf{e}_y + U_{dz}^{(0)} \mathbf{e}_z$, where $\mathbf{e}_x, \mathbf{e}_y$ and $\mathbf{e}_z$ are the unit vectors in *x*, *y* and *z* directions, respectively) to obtain the non-zero coefficients as (see appendix A)

$$\beta_{1,0}^{(0)} = -U_{dz}^{(0)}, \quad \beta_{1,1}^{(0)} = -U_{dx}^{(0)}, \quad \hat{\beta}_{1,1}^{(0)} = -U_{dy}^{(0)}, \tag{3.10}$$

where $U_{dx}^{(0)}, U_{dy}^{(0)}$ and $U_{dz}^{(0)}$ are the component of drop velocity at leading order in *x*,*y* and *z* directions, respectively.

The unknown coefficients $A_{n,m}^{(0)}, B_{n,m}^{(0)}, C_{n,m}^{(0)}, A_{-n-1,m}^{(0)}, B_{-n-1,m}^{(0)}$ and $C_{-n-1,m}^{(0)}$ can be determined by applying the boundary conditions given in equation (2.7). One thing to note here is that it is difficult to deal with the present form of the boundary conditions because some of the boundary conditions contain the derivative of solid spherical harmonics which makes the use of orthogonality property of spherical harmonics involved. This complexity can be avoided by presenting the boundary conditions in the following form (Hetsroni & Haber 1970; Happel & Brenner 1981)

$$\left[ \mathbf{u}_{i,r}^{(0)} \cdot \mathbf{e}_r \right]_s = 0,$$

$$\left[ \mathbf{v}_e^{(0)} \cdot \mathbf{e}_r \right]_s + \left[ \mathbf{u}_\infty^{(0)} \cdot \mathbf{e}_r \right]_s = 0,$$

$$\left[ r \frac{\partial}{\partial r} \left( \mathbf{u}_i^{(0)} \cdot \mathbf{e}_r \right) \right]_s = \left[ r \frac{\partial}{\partial r} \left( \mathbf{u}_\infty^{(0)} \cdot \mathbf{e}_r \right) \right] + \left[ r \frac{\partial}{\partial r} \left( \mathbf{v}_e^{(0)} \cdot \mathbf{e}_r \right) \right],$$

$$\left[ \mathbf{r} \cdot \nabla \times \mathbf{u}_i^{(0)} \right]_s = \left[ \mathbf{r} \cdot \nabla \times \mathbf{u}_\infty^{(0)} \right]_s + \left[ \mathbf{r} \cdot \nabla \times \mathbf{v}_e^{(0)} \right]_s,$$

$$\left[ \mathbf{r} \cdot \nabla \times \left\{ \mathbf{r} \times \left( \left( \boldsymbol{\tau}_i^{H(0)} + M \boldsymbol{\tau}_i^{E(0)} \right) \cdot \mathbf{e}_r \right) \right\} \right]_s = \left[ \mathbf{r} \cdot \nabla \times \left\{ \mathbf{r} \times \left( \boldsymbol{\tau}_\infty^{(0)} \cdot \mathbf{e}_r \right) \right\} \right]_s + \left[ \mathbf{r} \cdot \nabla \times \left\{ \mathbf{r} \times \left( \left( \boldsymbol{\tau}_e^{H(0)} + M \boldsymbol{\tau}_e^{E(0)} \right) \cdot \mathbf{e}_r \right) \right\} \right]_s,$$

$$\left[ \mathbf{r} \cdot \nabla \times \left\{ \left( \boldsymbol{\tau}_i^{H(0)} + M \boldsymbol{\tau}_i^{E(0)} \right) \cdot \mathbf{e}_r \right\} \right]_s = \left[ \mathbf{r} \cdot \nabla \times \left( \boldsymbol{\tau}_\infty^{(0)} \cdot \mathbf{e}_r \right) \right]_s + \left[ \mathbf{r} \cdot \nabla \times \left\{ \left( \boldsymbol{\tau}_e^{H(0)} + M \boldsymbol{\tau}_e^{E(0)} \right) \cdot \mathbf{e}_r \right\} \right]_s,$$

(3.11)

where $\left[ \Theta \right]_s$ represents the evaluation of a generic variable, $\Theta$, at the interface of the spherical drop $(r=1)$. The mathematical expressions of the hydrodynamic and electrical component of stresses are given by

$$\boldsymbol{\tau}_i^{H(0)} = \lambda \left[ -p_i^{(0)} \mathbf{I} + \nabla \mathbf{u}_i^{(0)} + \left( \nabla \mathbf{u}_i^{(0)} \right)^T \right],$$

$$\boldsymbol{\tau}_e^{H(0)} = \left[ -p_e^{(0)} \mathbf{I} + \nabla \mathbf{v}_e^{(0)} + \left( \nabla \mathbf{v}_e^{(0)} \right)^T \right],$$

$$\boldsymbol{\tau}_\infty^{H(0)} = \left[ -p_\infty^{(0)} \mathbf{I} + \nabla \mathbf{u}_\infty^{(0)} + \left( \nabla \mathbf{u}_\infty^{(0)} \right)^T \right], \quad (3.12)$$

$$\boldsymbol{\tau}_i^{E(0)} = S \left[ \mathbf{E}_i^{(0)} \left( \mathbf{E}_i^{(0)} \right)^T - \frac{1}{2} \left| \mathbf{E}_i^{(0)} \right|^2 \mathbf{I} \right],$$

$$\boldsymbol{\tau}_e^{E(0)} = \left[ \mathbf{E}_e^{(0)} \left( \mathbf{E}_e^{(0)} \right)^T - \frac{1}{2} \left| \mathbf{E}_e^{(0)} \right|^2 \mathbf{I} \right].$$

Applying these form of boundary conditions (see appendix B), we obtain the coefficients of the solid spherical harmonics in terms of the unknown drop velocity in the following form

$$A_{n,m}^{(0)} = \frac{(2n+3)\left[(2n+1)(2n-1)\beta_{n,m}^{(0)} + M\left(g_{n,m}^{i(0)} - g_{n,m}^{e(0)}\right)\right]}{n(2n+1)(\lambda+1)},$$

$$B_{n,m}^{(0)} = -\frac{A_{n,m}^{(0)}}{2(2n+3)},$$

$$C_{n,m}^{(0)} = \frac{M\left(h_{n,m}^{e(0)} - h_{n,m}^{i(0)}\right)}{n(n+1)(\lambda(n-1)+(n+2))},$$

$$A_{-n-1,m}^{(0)} = -\frac{\left[(4n^2-1)\{(2n+1)\lambda+2\}\beta_{n,m}^{(0)} + M(1-2n)\left(g_{n,m}^{i(0)} - g_{n,m}^{e(0)}\right)\right]}{(n+1)(2n+1)(\lambda+1)},$$

$$B_{-n-1,m}^{(0)} = -\frac{(4n^2-1)\lambda\beta_{n,m}^{(0)} + M\left(g_{n,m}^{e(0)} - g_{n,m}^{i(0)}\right)}{2(2n+1)(n+1)(\lambda+1)},$$

$$C_{-n-1,m}^{(0)} = C_{n,m}^{(0)},$$

(3.13)

keeping in mind that the following coefficients are non-zero from equation (3.10): $\beta_{1,0}^{(0)} = -U_{dz}^{(0)}, \beta_{1,1}^{(0)} = -U_{dx}^{(0)}$ and $\hat{\beta}_{1,1}^{(0)} = -U_{dy}^{(0)}$. Please note that the coefficients $g_{n,m}^{i,e(0)}, \hat{g}_{n,m}^{i,e(0)}, h_{n,m}^{i,e(0)}$ and $\hat{h}_{n,m}^{i,e(0)}$ are obtained from the surface harmonic representation of the electrical stresses as shown in Appendix B, equation (B5). Similar to the coefficients found in equation (3.13) the other coefficients $\hat{A}_{n,m}^{(0)}, \hat{B}_{n,m}^{(0)}, \hat{C}_{n,m}^{(0)}, \hat{A}_{-n-1,m}^{(0)}, \hat{B}_{-n-1,m}^{(0)}$ and $\hat{C}_{-n-1,m}^{(0)}$ are obtained by replacing $\beta_{n,m}^{(0)}, g_{n,m}^{i(0)}$ and $g_{n,m}^{e(0)}$ by $\hat{\beta}_{n,m}^{(0)}, \hat{g}_{n,m}^{i(0)}$ and $\hat{g}_{n,m}^{e(0)}$, respectively in equation (3.13). The complete expressions of velocity and pressure at the leading order are given in Appendix C.

### 3.2. *Leading order drop velocity: force free condition*

The sedimenting velocity of a drop is obtained by imposing the force free condition. At steady state the buoyancy force balances the hydrodynamic drag force on the drop. In a dimensional form this can be represented at the leading order as (Hetsroni & Haber 1970; Leal 2007)

$$\frac{4\pi}{3}a^3(\rho_i - \rho_e)g\mathbf{e}_z - 4\pi\nabla\left(r^3 p_{-2}^{(0)}\right) = 0. \quad (3.14)$$

Using the previously prescribed non-dimensional scheme equation (3.14) may be written as

$$\left(\frac{3\lambda+2}{\lambda+1}\right)\mathbf{e}_z - \nabla\left(r^3 p_{-2}^{(0)}\right) = 0, \quad (3.15)$$

where $\mathbf{e}_z = \cos\theta\mathbf{e}_r - \sin\theta\mathbf{e}_\theta$ and $p_{-2}^{(0)} = r^{-2}\left[A_{-2,0}^{(0)}P_{1,0} + \left(A_{-2,1}^{(0)}\cos\phi + \hat{A}_{-2,1}^{(0)}\sin\phi\right)P_{1,1}\right]$. Now, substituting $\mathbf{e}_z$ and $p_{-2}^{(0)}$ in equation (3.15), we obtain

| Non-zero coefficients of $\psi_i^{(1)}$ | Non-zero coefficients of $\psi_e^{(1)}$ |
|---|---|
| $a_{1,0}^{(1)}$ | $b_{-2,0}^{(1)}$ |
| $a_{2,0}^{(1)}, a_{2,1}^{(1)}, \hat{a}_{2,1}^{(1)}$ | $b_{-3,0}^{(1)}, b_{-3,1}^{(1)}, \hat{b}_{-3,1}^{(1)}$ |
| $a_{3,0}^{(1)}, a_{3,1}^{(1)}, \hat{a}_{3,1}^{(1)}, a_{3,2}^{(1)}, \hat{a}_{3,2}^{(1)}, a_{3,3}^{(1)}, \hat{a}_{3,3}^{(1)}$ | $b_{-4,0}^{(1)}, b_{-4,1}^{(1)}, \hat{b}_{-4,1}^{(1)}, b_{-4,2}^{(1)}, \hat{b}_{-4,2}^{(1)}, \hat{b}_{-4,3}^{(1)}, \hat{b}_{-4,3}^{(1)}$ |

TABLE 1. Non-zero coefficients present in first order electric potential.

$$A_{-2,0}^{(0)} = \frac{3\lambda + 2}{2(\lambda + 1)}, \quad A_{-2,1}^{(0)} = 0, \quad \hat{A}_{-2,1}^{(0)} = 0. \tag{3.16}$$

We obtain the drop velocity by using equation (3.16) and equation (3.13) of the form

$$U_{dx}^{(0)} = \frac{M\left(g_{1,1}^{e(0)} - g_{1,1}^{i(0)}\right)}{3(2+3\lambda)},$$

$$U_{dy}^{(0)} = \frac{M\left(\hat{g}_{1,1}^{e(0)} - \hat{g}_{1,1}^{i(0)}\right)}{3(2+3\lambda)}, \tag{3.17}$$

$$U_{dz}^{(0)} = 1 + \frac{M\left(g_{1,0}^{e(0)} - g_{1,0}^{i(0)}\right)}{3(2+3\lambda)}.$$

After solving leading order electric field we obtain $g_{n,m}^{i(0)} = g_{n,m}^{e(0)} = \hat{g}_{n,m}^{i(0)} = \hat{g}_{n,m}^{e(0)} = 0$, which gives the following drop velocity

$$U_{dx}^{(0)} = U_{dy}^{(0)} = 0, U_{dz}^{(0)} = 1. \tag{3.18}$$

### 3.3. *Second iteration - effect of surface convection on electric field*

Having obtained the leading order velocity profile from the previous subsection (details are given in Appendix C), we may now attempt the solution to equations (2.8) and (2.10) which are subjected to the boundary conditions (2.9) and (2.11), respectively. The first effect of charge convection appear in the charge convection boundary condition through the leading order velocity field and thus affects the first order electric field and, consequently, the first order velocity field. For convenience, we rewrite the boundary condition here: $\mathbf{e}_r \cdot \left(R\nabla \psi_i^{(1)} - \nabla \psi_e^{(1)}\right) = -\nabla_s \cdot \left(q^{(0)} \mathbf{V}_s^{(0)}\right)$. This differs from the leading order condition in the right hand side where it is zero, i.e. $\mathbf{e}_r \cdot \left(R\nabla \psi_i^{(0)} - \nabla \psi_e^{(0)}\right) = 0$.

Given that the potential fields are harmonic functions, the solutions are again given as

$$\psi_i^{(1)} = \sum_{n=0}^{\infty} \left( a_{n,m}^{(1)} \cos m\phi + \hat{a}_{n,m}^{(1)} \sin m\phi \right) r^n P_{n,m}(\cos\theta),$$
$$\psi_e^{(1)} = \sum_{n=0}^{\infty} \left( b_{n,m}^{(1)} \cos m\phi + \hat{b}_{n,m}^{(1)} \sin m\phi \right) r^{-n-1} P_{n,m}(\cos\theta). \tag{3.19}$$

Hence, it is instructive to first evaluate the right hand side of the charge convection boundary condition $\nabla_s \cdot \left( q^{(0)} \mathbf{V}_s^{(0)} \right)$ in order to establish the various surface harmonics that it may contain. It can be recast as (Kim & Karrila 1991)

$$-\nabla_s \cdot \left( q^{(0)} \mathbf{V}_s^{(0)} \right) = -2 \left( q^{(0)} \mathbf{V}_s^{(0)} \cdot \mathbf{e}_r \right) - \frac{1}{\sin\theta} \frac{\partial}{\partial \theta} \left( q^{(0)} \mathbf{V}_s^{(0)} \cdot \mathbf{e}_\theta \right) - \frac{1}{\sin\theta} \frac{\partial}{\partial \phi} \left( q^{(0)} \mathbf{V}_s^{(0)} \cdot \mathbf{e}_\phi \right). \tag{3.20}$$

By substituting the form of the velocity field (given in appendix C) and the accumulated charge (from equation (3.2)) in terms of the leading order electric field, the above expression can be split into various surface harmonics by making use of the orthogonality of the spherical surface harmonics. If we write expression (3.20) as $\sum_{n=0}^{\infty} \sum_{m=0}^{n} \left( Z_{n,m} \cos m\phi + \hat{Z}_{n,m} \sin m\phi \right) P_{n,m}$, we obtain the following nonzero terms: $Z_{1,0}, Z_{1,1}, Z_{2,0}, Z_{2,1}, Z_{3,0}, Z_{3,1}, Z_{3,2}, Z_{3,3}, \hat{Z}_{1,1}, \hat{Z}_{2,1}, \hat{Z}_{3,1}, \hat{Z}_{3,2}$ and $\hat{Z}_{3,3}$. The complete expressions of Zs are given in Appendix D. Thus it is expected that the first order electric field contains the non-zero coefficients $a_{n,m}^{(1)}, \hat{a}_{n,m}^{(1)}, b_{-n-1,m}^{(1)}$ and $\hat{b}_{-n-1,m}^{(1)}$ with $n$ and $m$ given in table 1. Now, we use rest of the boundary conditions from equation (2.9) and obtain the above unknown coefficients which are given in Appendix E.

### 3.4. $O(Re_E)$ drop velocity

The first order flow field also satisfies Stokes equation (equation (2.10)), so we can represent $\left( \mathbf{u}_i^{(1)}, p_i^{(1)} \right)$ and $\left( \mathbf{u}_e^{(1)}, p_e^{(1)} \right)$ in terms of the solid spherical harmonics $p_{n,m}^{(1)}, \Phi_{n,m}^{(1)}, \chi_{n,m}^{(1)}, p_{-n-1,m}^{(1)}, \Phi_{-n-1,m}^{(1)}$ and $\chi_{-n-1,m}^{(1)}$ using Lamb solution in a similar way as represented for the leading order solution in section 3.1. The details are worked out in Appendix F. In tune with the leading order drop velocity estimation, we may proceed in a similar fashion to evaluate the first order drop velocity

$$\mathbf{0} - 4\pi \nabla \left( r^3 p_{-2}^{(1)} \right) = 0. \tag{3.21}$$

Thus, employing the force-free condition at the first order, we obtain the following

$$A_{-2,0}^{(1)} = 0, A_{-2,1}^{(1)} = 0, \hat{A}_{-2,1}^{(1)} = 0. \tag{3.22}$$

This translates into the following higher order velocity:

$$U_{dx}^{(1)} = \frac{M\left(g_{1,1}^{e(1)} - g_{1,1}^{i(1)}\right)}{3(2+3\lambda)},$$

$$U_{dy}^{(1)} = \frac{M\left(\hat{g}_{1,1}^{e(1)} - \hat{g}_{1,1}^{i(1)}\right)}{3(2+3\lambda)}, \quad (3.23)$$

$$U_{dz}^{(1)} = \frac{M\left(g_{1,0}^{e(1)} - g_{1,0}^{i(1)}\right)}{3(2+3\lambda)}.$$

Substituting the following $g_{n,m}^{e,i(1)}$ and $\hat{g}_{n,m}^{e,i(1)}$ in equation (3.23), we obtain

$$U_{dx}^{(1)} = -\frac{3}{10}\frac{M(R-S)(3R-S+3)E_x E_z}{(3+2R)(R+2)^2(2+3\lambda)(\lambda+1)}, \quad (3.24)$$

$$U_{dy}^{(1)} = -\frac{3}{10}\frac{M(R-S)(3R-S+3)E_y E_z}{(3+2R)(R+2)^2(2+3\lambda)(\lambda+1)},$$

$$U_{dz}^{(1)} = -\frac{3}{10}\frac{M(R-S)(3R-S+3)\left(E_z^2+3\right)}{(3+2R)(R+2)^2(2+3\lambda)(\lambda+1)}. \quad (3.25)$$

Please note that we have made use of the identity $E_x^2 + E_y^2 + E_z^2 = 1$.

## 4. Summary and discussion

The general drop velocity can thus be written as the sum of the drop velocities obtained from the above two orders as found from equations (3.17) and (3.24). Mathematically,

$$\begin{aligned}\mathbf{U}_d &= \mathbf{U}_d^{(0)} + Re_E \mathbf{U}_d^{(1)} \\ &= \left[1 - Re_E \frac{3}{10}\frac{M(R-S)(3R-S+3)\left(E_z^2+3\right)}{(3+2R)(R+2)^2(2+3\lambda)(\lambda+1)}\right]\mathbf{e}_z \\ &\quad - \left[Re_E \frac{3}{10}\frac{M(R-S)(3R-S+3)E_x E_z}{(3+2R)(R+2)^2(2+3\lambda)(\lambda+1)}\right]\mathbf{e}_x \\ &\quad - \left[Re_E \frac{3}{10}\frac{M(R-S)(3R-S+3)E_y E_z}{(3+2R)(R+2)^2(2+3\lambda)(\lambda+1)}\right]\mathbf{e}_y\end{aligned}$$

(4.1)

If we consider an applied electric field only in the z-direction, i.e. $\mathbf{E}_\infty = (0,0,1)$, we obtain

$$\mathbf{U}_d = \left[1 - Re_E \frac{6}{5} \frac{M(R-S)(3R-S+3)}{(3+2R)(R+2)^2(2+3\lambda)(\lambda+1)}\right] \mathbf{e}_z, \quad (4.2)$$

which corroborates with the drop sedimentation velocity obtained by Xu and Homsy (2006).

It is interesting to note one particular aspect of the drop velocity obtained above. When an electric field is applied only in the *x*-direction, we still get a change in the sedimentation velocity while there is no lateral migration velocity. However, when the applied electric field has all three components as non-zero, i.e. $\mathbf{E}_\infty = (E_x, E_y, E_z)$, not only does the sedimentation get affected but there is also a lateral migration observed in both the *x* and *y* directions. Towards discussing the physical reason behind the observations, we consider the following three cases: (a) $\mathbf{E}_\infty = (0,0,1)$ (longitudinal electric field), (b) $\mathbf{E}_\infty = (1,0,0)$ (transverse electric field) and (c) $\mathbf{E}_\infty = (1/\sqrt{2}, 0, 1/\sqrt{2})$ (both longitudinal and transverse fields).

### 4.1. *Longitudinal electric field*

Under the assumption of a neutrally buoyant droplet, advancing the problem considered by Taylor (1966), Feng (1999) showed that the consideration of surface charge convection led to alterations in the drop deformation. It is to be noted that despite charge convection, for a neutrally buoyant drop, the charge distribution remains anti-symmetric along the equatorial plane. The consequence of this is that there is no extra drag acting on the drop and consequently the drop remains stationary. The problem of a drop with a different density than the suspending fluid in the presence of longitudinal applied electric field was then considered by Xu and Homsy (2006) who highlighted the pivotal role of the surface charge convection in breaking the anti-symmetry about the equatorial plane thus giving rise to a net drag on the drop. This can be seen from figure 2 which is derived using the expressions in section 3 and setting $E_z = 1$ while setting the other components of the electric field to be zero. The choice of the parameters used in figure 2 is mentioned in the figure caption. Figure 2(a) depicts the leading order charge distribution which is anti-symmetric about the equatorial plane. However, it can be clearly seen from figure 2(b) that the surface convection for a sedimenting drop *does* lead to a breaking in the anti-symmetric structure of the charge distribution. It is precisely this asymmetric charge distribution that results in the alteration in the sedimentation velocity.

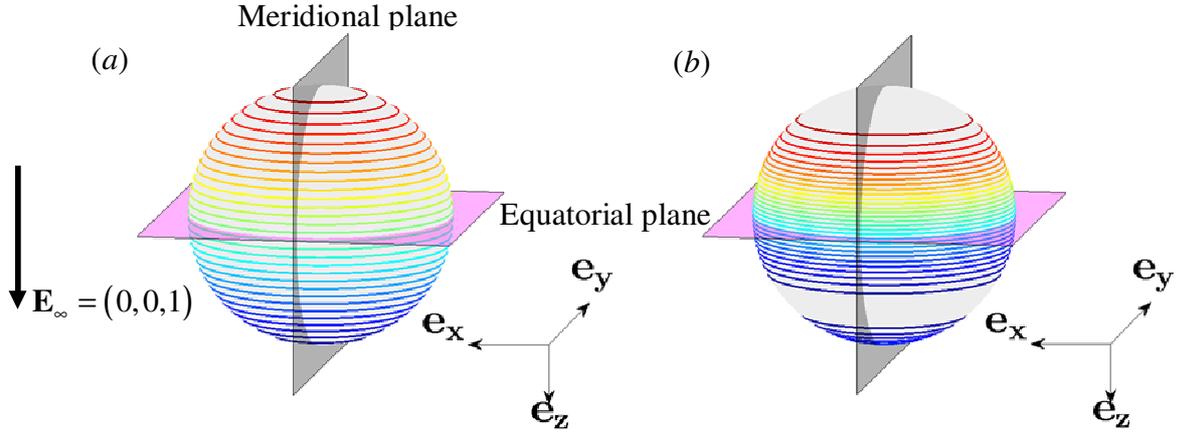

FIGURE 2. (Colour online) Contours depicting the variation of the (a) $q^{(0)}$ and (b) $q^{(0)} + Re_E q^{(1)}$ for a longitudinal electric field (case I). The parameters employed are $R = 0.5$, $S = 2$, $M = 2$, $\lambda = 0.5$ and $Re_E = 0.2$.

### 4.2. *Transverse electric field*

A striking result obtained in equation (4.1) is that even if there is an applied electric field in transverse direction, the influence of charge convection is solely affecting the sedimentation velocity without causing any migration in the lateral direction. To understand this from the view point of symmetries in the charge distribution, we appeal to figure 3. It is seen from figure 3(a) that the absence of surface convection leads to an anti-symmetric structure about the meridional plane and maintains a symmetry about the equatorial plane. In the presence of surface convection due to sedimentation, however, it is observed that the anti-symmetric structure is preserved about the meridional plane, but the symmetry is broken about the equatorial plane. This leads to a net force in the longitudinal direction without affecting the motion in the transverse direction, leading to an alteration in the sedimentation velocity.

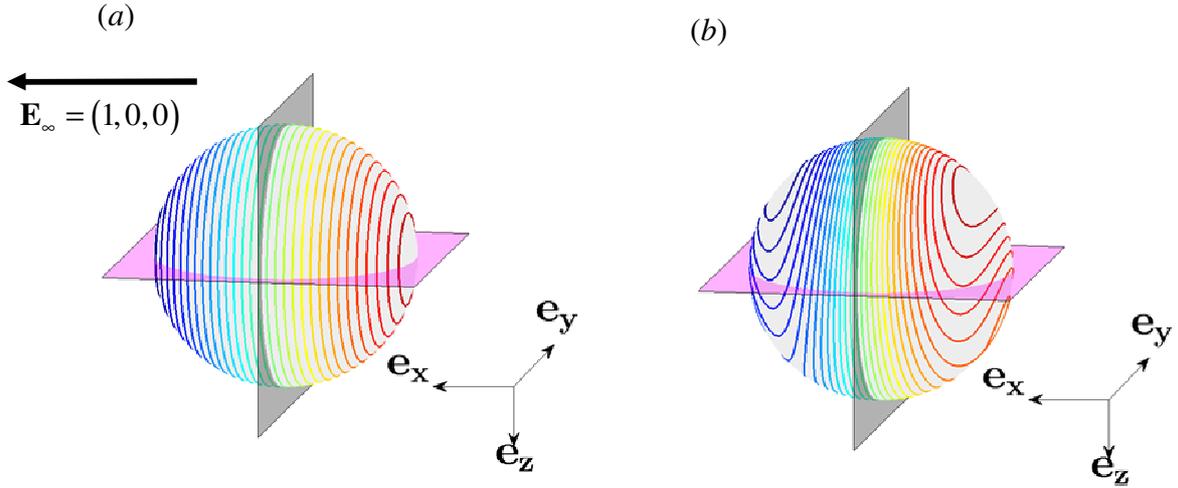

FIGURE 3. (Colour online) Contours depicting the variation of the (a) $q^{(0)}$ and (b) $q^{(0)} + Re_E q^{(1)}$ for a transverse electric field (case II). The parameters employed are $R = 0.5$, $S = 2$, $M = 2$, $\lambda = 0.5$ and $Re_E = 0.2$.

### 4.3. *Combined longitudinal and transverse electric field*

We depict the charge distribution due to surface convection in figure 4 where it can be seen that the combined nature of the applied electric field causes an asymmetry about both the equatorial and meridional plane leading to a force in both the longitudinal and transverse directions respectively. The findings are also seen in the expressions for the droplet velocity by setting $\mathbf{E}_\infty = \left(1/\sqrt{2}, 0, 1/\sqrt{2}\right)$ in equation (4.1). A transverse or longitudinal applied electric field alone is insufficient to cause any lateral migration.

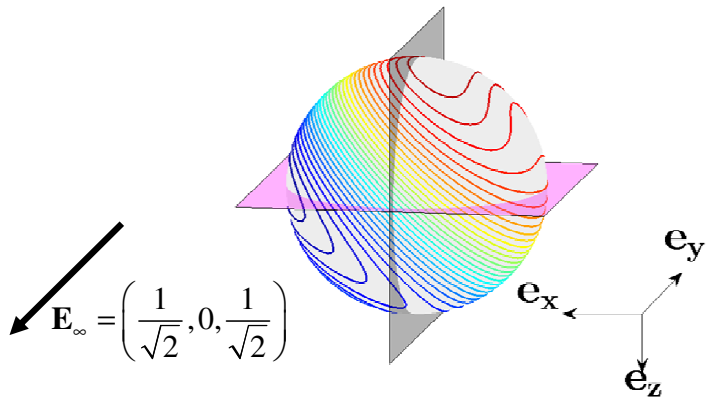

FIGURE 4. (Colour online) Contours depicting the variation of $q^{(0)} + Re_E q^{(1)}$ for a combined longitudinal and transverse electric field (case III). The parameters employed are $R = 0.5$, $S = 2$, $M = 2$, $\lambda = 0.5$ and $Re_E = 0.2$.

## 5. Conclusions

We conclude by remarking on the direction of lateral velocity of drop in the presence of both longitudinal and transverse electric field. Depending on the combination of the direction of applied electric field and the ratios of permittivities and conductivities, the direction of the motion is decided. Towards this, figure 5 depicts the two regimes of the transverse velocity of drop in $x$ direction $(U_{dx})$ in the $R-S$ plane considering $\mathbf{E}_\infty = (1/\sqrt{2}, 0, 1/\sqrt{2})$. We identify two regime boundaries given by $R = S$ and $S = 3R + 3$, where the former case is tantamount to a perfect dielectric case with zero circulation while the latter case denotes the zero-lateral-migration line for a leaky dielectric drop (in accordance with equation (4.1) for $U_{dx} = 0$).

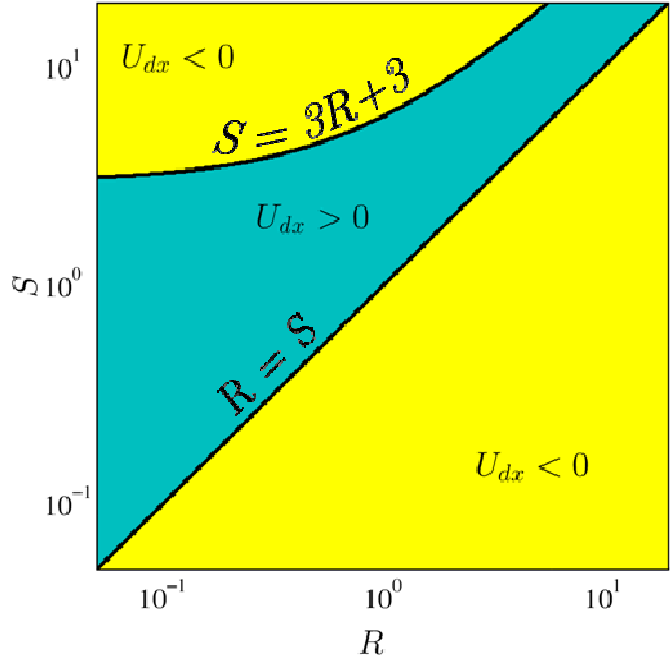

FIGURE 5. (Colour online) Different regimes of lateral velocity of drop $(U_{dx})$ in $R-S$ plane. The parameters employed are $M = 2$, $\lambda = 0.5$ and $Re_E = 0.2$.

It is to be noted that in the above case, the lateral migration depends on the product of $E_x$ and $E_z$. As a result, the migration velocity, $U_{dx}$, remains unaltered by the reversal of direction of $\mathbf{E}_\infty$. In conclusion, we have demonstrated that the arbitrary nature of the uniform electric field is responsible for the lateral migration of an otherwise longitudinally sedimenting drop for small $Re_E$. The asymmetry in the charge distribution about the meridional plane is caused due to the simultaneous action of the electric field in both the longitudinal and transverse directions. The impending sedimentation velocity is attributed to the tendency of the drop to migrate laterally while sedimenting. The analytical results presented here are expected to have a broad appeal to scientists and engineers alike who are concerned with processes pertaining to electrohydrodynamic drop motion and manipulation.

**Appendix A. Derivation of the relation between the coefficients present in the solid spherical harmonics present in $\mathbf{u}_\infty^{(0)}$ with the component of drop velocity $\mathbf{U}_d^{(0)}$**

The coefficients present in $\mathbf{u}_\infty^{(0)}$ can be related to the cartesian components of drop velocity $U_{dx}^{(0)}, U_{dy}^{(0)}$ and $U_{dz}^{(0)}$. At first we obtain the dot product of unit radial vector with $\mathbf{u}_\infty^{(0)}$ as (Hetsroni & Haber 1970)

$$\mathbf{u}_\infty^{(0)} \cdot \mathbf{e}_r = \sum_{n=1}^{\infty} \sum_{m=0}^{n} \left[ \alpha_{n,m}^{(0)} r^{n+1} + \beta_{n,m}^{(0)} r^{n-1} + \alpha_{-n-1,m}^{(0)} r^{-n} + \beta_{-n-1,m}^{(0)} r^{-n-2} \right] P_{n,m} \cos m\phi$$
$$+ \sum_{n=1}^{\infty} \sum_{m=0}^{n} \left[ \hat{\alpha}_{n,m}^{(0)} r^{n+1} + \hat{\beta}_{n,m}^{(0)} r^{n-1} + \hat{\alpha}_{-n-1,m}^{(0)} r^{-n} + \hat{\beta}_{-n-1,m}^{(0)} r^{-n-2} \right] P_{n,m} \sin m\phi. \quad (A1)$$

As $\mathbf{u}_\infty^{(0)} = -\mathbf{U}_d^{(0)}$, by taking the dot product of both the sides with unit radial vector and using equation (A1), we obtain

$$\sum_{n=1}^{\infty} \sum_{m=0}^{n} \left[ \alpha_{n,m}^{(0)} r^{n+1} + \beta_{n,m}^{(0)} r^{n-1} + \alpha_{-n-1,m}^{(0)} r^{-n} + \beta_{-n-1,m}^{(0)} r^{-n-2} \right] P_{n,m} \cos m\phi$$
$$+ \sum_{n=1}^{\infty} \sum_{m=0}^{n} \left[ \hat{\alpha}_{n,m}^{(0)} r^{n+1} + \hat{\beta}_{n,m}^{(0)} r^{n-1} + \hat{\alpha}_{-n-1,m}^{(0)} r^{-n} + \hat{\beta}_{-n-1,m}^{(0)} r^{-n-2} \right] P_{n,m} \sin m\phi \quad (A2)$$
$$= -\left[ U_{dx}^{(0)} \cos\phi P_{1,1} + U_{dy}^{(0)} \sin\phi P_{1,1} + U_{dz}^{(0)} P_{1,0} \right].$$

Now, comparing both sides of equation (A2) we obtain the non-zero $\alpha$ and $\beta$ in the following form

$$\beta_{1,0}^{(0)} = -U_{dz}^{(0)}, \quad \beta_{1,1}^{(0)} = -U_{dx}^{(0)}, \quad \hat{\beta}_{1,1}^{(0)} = -U_{dy}^{(0)}. \quad (A3)$$

For other values of $n$ and $m$, the coefficients $\alpha^{(0)}$ and $\beta^{(0)}$ vanish. The coefficient $\gamma^{(0)}$ can be obtained by taking $\mathbf{r} \cdot \nabla \times$ operation over both sides of $\mathbf{u}_\infty^{(0)} = -\mathbf{U}_d^{(0)}$. We obtain

$$\mathbf{r} \cdot \nabla \times \mathbf{u}_\infty^{(0)} = \mathbf{r} \cdot \nabla \times \mathbf{U}_d^{(0)} = 0, \quad (A4)$$

where $\mathbf{r}\cdot\nabla\times\mathbf{u}_\infty^{(0)}$ can be expressed as (Hetsroni & Haber 1970)

$$\mathbf{r}\cdot\nabla\times\mathbf{u}_\infty^{(0)} = \sum_{n=1}^{\infty}\sum_{m=0}^{n}\left[\left\{\gamma_{n,m}^{(0)}r^n + \gamma_{-n-1,m}^{(0)}r^{-n-1}\right\}\cos m\phi + \left\{\hat{\gamma}_{n,m}^{(0)}r^n + \hat{\gamma}_{-n-1,m}^{(0)}r^{-n-1}\right\}\sin m\phi\right]P_{n,m}. \quad (A5)$$

Now, comparing equations (A4) and (A5), we obtain $\gamma_{n,m}^{(0)} = \hat{\gamma}_{n,m}^{(0)} = \gamma_{-n-1,m}^{(0)} = \hat{\gamma}_{-n-1,m}^{(0)} = 0$ for all values of $n$ and $m$. Similar relations can also be obtained for $O(Re_E)$.

**Appendix B. Implementation of boundary conditions to obtain the solid spherical harmonics present in the Lamb solution**

The implementation of the velocity and stress boundary conditions are summarized here. To apply the boundary conditions given in equation (3.11), at first one has to express $\left[\mathbf{u}^{(0)}\cdot\mathbf{e}_r\right]_s$, $\left[r\dfrac{\partial}{\partial r}\left(\mathbf{u}^{(0)}\cdot\mathbf{e}_r\right)\right]_s$ and $\left[\mathbf{r}\cdot\nabla\times\mathbf{u}^{(0)}\right]_s$ in terms of solid spherical harmonics in the following form (Happel & Brenner 1981)

$$\begin{aligned}
\left[\mathbf{u}^{(0)}\cdot\mathbf{e}_r\right]_s &= \sum_{n=-\infty}^{\infty}\left[\frac{n}{2N_\mu(2n+3)}p_n^{(0)} + n\Phi_n^{(0)}\right], \\
\left[r\frac{\partial}{\partial r}\left(\mathbf{u}^{(0)}\cdot\mathbf{e}_r\right)\right]_s &= \sum_{n=-\infty}^{\infty}\left[\frac{n(n+1)}{2N_\mu(2n+3)}p_n^{(0)} + n(n-1)\Phi_n^{(0)}\right], \quad (B1) \\
\left[\mathbf{r}\cdot\nabla\times\mathbf{u}^{(0)}\right]_s &= \sum_{n=-\infty}^{\infty}n(n+1)\chi_n^{(0)}.
\end{aligned}$$

To satisfy the shear stress balance, we have to obtain $\left[\mathbf{r}\cdot\nabla\times\left\{\mathbf{r}\times\left(\left(\boldsymbol{\tau}^{H(0)} + M\boldsymbol{\tau}^{E(0)}\right)\cdot\mathbf{e}_r\right)\right\}\right]_s$ and $\left[\mathbf{r}\cdot\nabla\times\left\{\left(\boldsymbol{\tau}^{H(0)} + M\boldsymbol{\tau}^{E(0)}\right)\cdot\mathbf{e}_r\right\}\right]_s$. Among these terms the hydrodynamic stress components can be obtained as

$$\left[\mathbf{r}\cdot\nabla\times\left(\mathbf{r}\times\left(\boldsymbol{\tau}^{H(0)}\cdot\mathbf{e}_r\right)\right)\right]_s = -N_\mu\sum_{n=-\infty}^{\infty}\left[2(n-1)n(n+1)\Phi_n^{(0)} + \frac{n^2(n+2)}{N_\mu(2n+3)}p_n^{(0)}\right], \quad (B2)$$

$$\left[\mathbf{r}\cdot\nabla\times\left(\boldsymbol{\tau}^{H(0)}\cdot\mathbf{e}_r\right)\right]_s = N_\mu\sum_{n=-\infty}^{\infty}(n-1)n(n+1)\chi_n^{(0)}. \quad (B3)$$

The radial traction vector, $\boldsymbol{\tau}_i^{E(0)}\cdot\mathbf{e}_r$ and $\boldsymbol{\tau}_e^{E(0)}\cdot\mathbf{e}_r$, arising from the Maxwell stress tensor arising due to the zeroth order electric field are expressed as

$$\boldsymbol{\tau}_i^{E(0)} \cdot \mathbf{e}_r = S \begin{bmatrix} \left(E_{i,r}^{(0)}\right)^2 - \frac{1}{2}\left|\mathbf{E}_{i,r}^{(0)}\right|^2 \\ E_{i,r}^{(0)} E_{i,\theta}^{(0)} \\ E_{i,r}^{(0)} E_{i,\phi}^{(0)} \end{bmatrix}, \quad \boldsymbol{\tau}_e^{E(0)} \cdot \mathbf{e}_r = \begin{bmatrix} \left(E_{e,r}^{(0)}\right)^2 - \frac{1}{2}\left|\mathbf{E}_{e,r}^{(0)}\right|^2 \\ E_{e,r}^{(0)} E_{e,\theta}^{(0)} \\ E_{e,r}^{(0)} E_{e,\phi}^{(0)} \end{bmatrix}. \tag{B4}$$

We express the electrical stress components at the drop interface in terms of surface harmonics as

$$\left[\mathbf{r} \cdot \nabla \times \left\{\mathbf{r} \times \left(\boldsymbol{\tau}_i^{E(0)} \cdot \mathbf{e}_r\right)\right\}\right]_s = \sum_{n=0}^{\infty} \left(g_{n,m}^{i(0)} \cos m\phi + \hat{g}_{n,m}^{i(0)} \sin m\phi\right) P_{n,m},$$

$$\left[\mathbf{r} \cdot \nabla \times \left\{\mathbf{r} \times \left(\boldsymbol{\tau}_i^{E(0)} \cdot \mathbf{e}_r\right)\right\}\right]_s = \sum_{n=0}^{\infty} \left(g_{n,m}^{e(0)} \cos m\phi + \hat{g}_{n,m}^{e(0)} \sin m\phi\right) P_{n,m},$$

$$\left[\mathbf{r} \cdot \nabla \times \left(\boldsymbol{\tau}_i^{E(0)} \cdot \mathbf{e}_r\right)\right]_s = \sum_{n=0}^{\infty} \left(h_{n,m}^{i(0)} \cos m\phi + \hat{h}_{n,m}^{i(0)} \sin m\phi\right) P_{n,m},$$

$$\left[\mathbf{r} \cdot \nabla \times \left(\boldsymbol{\tau}_e^{E(0)} \cdot \mathbf{e}_r\right)\right]_s = \sum_{n=0}^{\infty} \left(h_{n,m}^{e(0)} \cos m\phi + \hat{h}_{n,m}^{e(0)} \sin m\phi\right) P_{n,m},$$

(B5)

where $g_{n,m}^{i(0)}, \hat{g}_{n,m}^{i(0)}, h_{n,m}^{i(0)}, \hat{h}_{n,m}^{i(0)}, g_{n,m}^{e(0)}, \hat{g}_{n,m}^{e(0)}, h_{n,m}^{e(0)}$ and $\hat{h}_{n,m}^{e(0)}$ are the coefficients that can be obtained after solving electrical potential distribution. The nonzero coefficients obtained from the leading order electric potential (equation (3.2)) are:

$$g_{2,0}^{i(0)} = \frac{9\left(E_y^2 + E_x^2 - 2E_z^2\right)S}{(R+2)^2}, \quad g_{2,1}^{i(0)} = -\frac{18SE_z E_x}{(R+2)^2}, \quad g_{2,2}^{i(0)} = -\frac{9}{2}\frac{\left(E_x^2 - E_y^2\right)S}{(R+2)^2},$$

$$\hat{g}_{2,1}^{i(0)} = \frac{-18 E_y E_z S}{(R+2)^2}, \quad \hat{g}_{2,2}^{i(0)} = -\frac{9 E_x E_y S}{(R+2)^2}, \quad g_{2,0}^{e(0)} = \frac{9\left(E_x^2 + E_y^2 - 2E_z^2\right)R}{(R+2)^2},$$

$$g_{2,1}^{e(0)} = \frac{-18 R E_z E_x}{(R+2)^2}, \quad g_{2,2}^{e(0)} = -\frac{9}{2}\frac{\left(-E_y^2 + E_x^2\right)R}{(R+2)^2}, \quad \hat{g}_{2,1}^{e(0)} = \frac{-18 E_y E_z R}{(R+2)^2},$$

$$\hat{g}_{2,2}^{e(0)} = -\frac{9 E_x E_y R}{(R+2)^2}.$$

(B6)

**Appendix C. Velocity and pressure field at leading order**

The velocity and pressure fields inside the drop is given by

$$\mathbf{u}_i^{(0)} = \left[\nabla\left(\Phi_1^{(0)} + \Phi_2^{(0)}\right) + r^2 \nabla\left(\frac{1}{5\lambda}p_1^{(0)} + \frac{5}{42\lambda}p_2^{(0)}\right) - \mathbf{r}\left(\frac{1}{10\lambda}p_1^{(0)} + \frac{2}{21\lambda}p_2^{(0)}\right)\right],$$

$$p_i^{(0)} = p_1^{(0)} + p_2^{(0)},$$

(C1)

where the solid harmonics are of the following form

$$\Phi_1^{(0)} = rB_{1,0}^{(0)}P_{1,0},$$

$$\Phi_2^{(0)} = r^2 \left[ B_{2,0}^{(0)} P_{2,0} + \left( B_{2,1}^{(0)} \cos\phi + \hat{B}_{2,1}^{(0)} \sin\phi \right) P_{2,1} + \left( B_{2,2}^{(0)} \cos 2\phi + \hat{B}_{2,2}^{(0)} \sin 2\phi \right) P_{2,2} \right],$$

$$p_1^{(0)} = \lambda r A_{1,0}^{(0)} P_{1,0},$$

$$p_2^{(0)} = \lambda r^2 \left[ A_{2,0}^{(0)} P_{2,0} + \left( A_{2,1}^{(0)} \cos\phi + \hat{A}_{2,1}^{(0)} \sin\phi \right) P_{2,1} + \left( A_{2,2}^{(0)} \cos 2\phi + \hat{A}_{2,2}^{(0)} \sin 2\phi \right) P_{2,2} \right]. \quad \text{(C2)}$$

The coefficients $A^{(0)}$, $\hat{A}^{(0)}$, $B^{(0)}$ and $\hat{B}^{(0)}$ present in the above solid harmonics can be obtained using equation (3.13) of the following form

$$A_{1,0}^{(0)} = -\frac{5}{\lambda+1}, \ A_{2,0}^{(0)} = -\frac{63}{5}\left(1-3E_z^2\right)\Lambda, \ A_{2,1}^{(0)} = \frac{63}{5} E_x E_z \Lambda, \ A_{2,2}^{(0)} = \frac{63}{20}\left(E_x^2 - E_y^2\right)\Lambda,$$

$$\hat{A}_{2,1}^{(0)} = \frac{63}{5} E_y E_z \Lambda, \ \hat{A}_{2,2}^{(0)} = \frac{63}{10} E_x E_y \Lambda, \ B_{1,0}^{(0)} = \frac{1}{2(\lambda+1)}, \ B_{2,0}^{(0)} = \frac{9}{20}\left(1-3E_z^2\right)\Lambda, \quad \text{(C3)}$$

$$B_{2,1}^{(0)} = -\frac{9}{10} E_x E_z \Lambda, \ B_{2,2}^{(0)} = -\frac{9}{40}\left(E_x^2 - E_y^2\right)\Lambda, \ \hat{B}_{2,1}^{(0)} = -\frac{9}{10} E_y E_z \Lambda, \ \hat{B}_{2,2}^{(0)} = -\frac{9}{20} E_x E_y \Lambda,$$

where the parameter $\Lambda = \dfrac{M(R-S)}{(\lambda+1)(R+2)^2}$.

Similarly, the velocity and pressure fields external to the drop are given by

$$\mathbf{u}_e^{(0)} = -\mathbf{U}_d^{(0)} + \mathbf{v}_e^{(0)}$$

$$= -\left(\cos\theta \mathbf{e}_r - \sin\theta \mathbf{e}_\theta\right) + \sum_{n=1}^{\infty} \left[ \nabla\left(\Phi_{-2}^{(0)} + \Phi_{-3}^{(0)}\right) - \frac{1}{2} r^2 \nabla p_{-2}^{(0)} + \mathbf{r}\left( 2 p_{-2}^{(0)} + \frac{1}{2} p_{-3}^{(0)} \right) \right], \quad \text{(C4)}$$

$$p_e^{(0)} = p_{-2}^{(0)} + p_{-3}^{(0)},$$

The solid harmonics are of the following form

$$\Phi_{-2}^{(0)} = \frac{1}{r^2} B_{-2,0}^{(0)} P_{1,0}, \ p_{-2}^{(0)} = \frac{1}{r^2} A_{-2,0}^{(0)} P_{1,0},$$

$$\Phi_{-3}^{(0)} = \frac{1}{r^3}\left[ B_{-3,0}^{(0)} P_{2,0} + \left( B_{-3,1}^{(0)} \cos\phi + \hat{B}_{-3,1}^{(0)} \sin\phi \right) P_{2,1} + \left( B_{-3,2}^{(0)} \cos 2\phi + \hat{B}_{-3,2}^{(0)} \sin 2\phi \right) P_{2,2} \right], \quad \text{(C5)}$$

$$p_{-3}^{(0)} = \frac{1}{r^3}\left[ A_{-3,0}^{(0)} P_{2,0} + \left( A_{-3,1}^{(0)} \cos\phi + \hat{A}_{-3,1}^{(0)} \sin\phi \right) P_{2,1} + \left( A_{-3,2}^{(0)} \cos 2\phi + \hat{A}_{-3,2}^{(0)} \sin 2\phi \right) P_{2,2} \right].$$

The coefficients $A^{(0)}$, $\hat{A}^{(0)}$, $B^{(0)}$ and $\hat{B}^{(0)}$ present in the above solid harmonics can be obtained using equation (3.13) of the following form

$$A_{-2,0}^{(0)} = \frac{2+3\lambda}{2(\lambda+1)}, \quad A_{-3,0}^{(0)} = -\frac{9}{5}(1-3E_z^2)\Lambda, \quad A_{-3,1}^{(0)} = \frac{18}{5}E_x E\Lambda, \quad A_{-3,2}^{(0)} = \frac{9}{10}(E_x^2 - E_y^2)\Lambda,$$

$$\hat{A}_{-3,1}^{(0)} = \frac{18}{5}E_y E_z\Lambda, \quad \hat{A}_{-3,2}^{(0)} = \frac{9}{5}E_x E_y\Lambda, \quad B_{-2,0}^{(0)} = \frac{\lambda}{4(\lambda+1)}, \quad B_{-3,0}^{(0)} = -\frac{3}{10}(1-3E_z^2)\Lambda, \quad (C6)$$

$$B_{-3,1}^{(0)} = \frac{3}{5}E_x E_z\Lambda, \quad B_{-3,2}^{(0)} = \frac{3}{20}(E_x^2 - E_y^2)\Lambda, \quad \hat{B}_{-3,1}^{(0)} = \frac{3}{5}E_y E_z\Lambda, \quad \hat{B}_{-3,2}^{(0)} = \frac{3}{10}E_x E_y\Lambda.$$

**Appendix D. Complete expressions of Zs present at the right hand side of charge conservation equation at $O(Re_E)$**

The complete expressions of the non-zero $Z_{n,m}$ are given as

$$Z_{1,0} = \frac{54}{25}\frac{M(R-S)^2 E_z}{(R+2)^3(\lambda+1)}, \quad Z_{1,1} = \frac{54}{25}\frac{M(R-S)^2 E_x}{(R+2)^3(\lambda+1)}, \quad \hat{Z}_{1,1} = \frac{54}{25}\frac{M(R-S)^2 E_y}{(R+2)^3(\lambda+1)},$$

$$Z_{2,0} = -\frac{3(R-S)E_z}{(R+2)(\lambda+1)}, \quad Z_{2,1} = -\frac{3}{2}\frac{(R-S)E_x}{(R+2)(\lambda+1)}, \quad \hat{Z}_{2,1} = -\frac{3}{2}\frac{(R-S)E_y}{(R+2)(\lambda+1)},$$

$$Z_{3,0} = -\frac{216}{25}\frac{M(R-S)^2 E_z}{(R+2)^3(\lambda+1)}, \quad Z_{3,1} = -\frac{216}{25}\frac{M(R-S)^2 E_x}{(R+2)^3(\lambda+1)}, \quad \hat{Z}_{3,1} = -\frac{216}{25}\frac{M(R-S)^2 E_y}{(R+2)^3(\lambda+1)},$$

$$Z_{3,2} = \frac{54}{25}\frac{M(R-S)^2 E_z(E_x^2 - E_y^2)}{(R+2)^3(\lambda+1)}, \quad \hat{Z}_{3,2} = \frac{108}{25}\frac{M(R-S)^2 E_x E_y E_z}{(R+2)^3(\lambda+1)},$$

$$Z_{3,3} = \frac{9}{25}\frac{M(R-S)^2 E_x(E_x^2 - 3E_y^2)}{(R+2)^3(\lambda+1)}, \quad \hat{Z}_{3,3} = \frac{9}{25}\frac{M(R-S)^2 E_y(3E_x^2 - E_y^2)}{(R+2)^3(\lambda+1)}.$$

(D1)

**Appendix E. The $O(Re_E)$ electric field**

The electric potential at $O(Re_E)$ is given by

$$\psi_i^{(1)} = r\left[a_{1,0}^{(1)}P_{1,0} + \left(a_{1,1}^{(1)}\cos\phi + \hat{a}_{1,1}^{(1)}\sin\phi\right)P_{1,1}\right] + r^2\left[a_{2,0}^{(1)}P_{2,0} + \left(a_{2,1}^{(1)}\cos\phi + \hat{a}_{2,1}^{(1)}\sin\phi\right)P_{2,1}\right]$$

$$+ r^3\left[a_{3,0}^{(1)}P_{3,0} + \left(a_{3,1}^{(1)}\cos\phi + \hat{a}_{3,1}^{(1)}\sin\phi\right)P_{3,1} + \left(a_{3,2}^{(1)}\cos 2\phi + \hat{a}_{3,2}^{(1)}\sin 2\phi\right)P_{3,2}\right.$$

$$\left. + \left(a_{3,3}^{(1)}\cos 3\phi + \hat{a}_{3,3}^{(1)}\sin 3\phi\right)P_{3,3}\right],$$

(E1)

$$\psi_e^{(1)} = \frac{1}{r^2}\left[b_{-2,0}^{(1)} P_{1,0} + \left(b_{-2,1}^{(1)} \cos\phi + \hat{b}_{-2,1}^{(1)} \sin\phi\right) P_{1,1}\right] + \frac{1}{r^3}\left[b_{-3,0}^{(1)} P_{2,0} + \left(b_{-3,1}^{(1)} \cos\phi + \hat{b}_{-3,1}^{(1)} \sin\phi\right) P_{2,1}\right]$$
$$+ \frac{1}{r^4}\left[b_{-4,0}^{(1)} P_{3,0} + \left(b_{-4,1}^{(1)} \cos\phi + \hat{b}_{-4,1}^{(1)} \sin\phi\right) P_{3,1} + \left(b_{-4,2}^{(1)} \cos 2\phi + \hat{b}_{-4,2}^{(1)} \sin 2\phi\right) P_{3,2}\right.$$
$$\left. + \left(b_{-4,3}^{(1)} \cos 3\phi + \hat{b}_{-4,3}^{(1)} \sin 3\phi\right) P_{3,3}\right],$$

(E2)

where the non-zero coefficients present in the above expressions can be given in terms of $Z_{n,m}$ in the following form

$$a_{1,0}^{(1)} = \frac{Z_{1,0}}{2+R},\ a_{1,1}^{(1)} = \frac{Z_{1,1}}{2+R},\ a_{2,0}^{(1)} = \frac{Z_{2,0}}{3+2R},\ a_{2,1}^{(1)} = \frac{Z_{2,1}}{3+2R},\ a_{3,0}^{(1)} = \frac{Z_{3,0}}{4+3R},\ a_{3,1}^{(1)} = \frac{Z_{3,1}}{4+3R},$$
$$a_{3,2}^{(1)} = \frac{Z_{3,2}}{4+3R},\ a_{3,3}^{(1)} = \frac{Z_{3,3}}{4+3R},\ \hat{a}_{1,1}^{(1)} = \frac{\hat{Z}_{1,1}}{2+R},\ \hat{a}_{2,1}^{(1)} = \frac{\hat{Z}_{2,1}}{3+2R},\ \hat{a}_{3,1}^{(1)} = \frac{\hat{Z}_{3,1}}{4+3R},\ \hat{a}_{3,2}^{(1)} = \frac{\hat{Z}_{3,2}}{4+3R},$$
$$\hat{a}_{3,3}^{(1)} = \frac{\hat{Z}_{3,3}}{4+3R},\ b_{-2,0}^{(1)} = a_{1,0}^{(1)},\ b_{-2,1}^{(1)} = a_{1,1}^{(1)},\ b_{-3,0}^{(1)} = a_{2,0}^{(1)},\ b_{-3,1}^{(1)} = a_{2,1}^{(1)},\ b_{-4,0}^{(1)} = a_{3,0}^{(1)},\ b_{-4,1}^{(1)} = a_{3,1}^{(1)},$$
$$b_{-4,2}^{(1)} = a_{3,2}^{(1)},\ b_{-4,3}^{(1)} = a_{3,3}^{(1)},\ \hat{b}_{-2,1}^{(1)} = \hat{a}_{1,1}^{(1)},\ \hat{b}_{-3,1}^{(1)} = \hat{a}_{2,1}^{(1)},\ \hat{b}_{-4,1}^{(1)} = \hat{a}_{3,1}^{(1)},\ \hat{b}_{-4,2}^{(1)} = \hat{a}_{3,2}^{(1)},\ \hat{b}_{-4,3}^{(1)} = \hat{a}_{3,3}^{(1)}.$$

(E3)

## Appendix F. Non-zero solid harmonics of $O(Re_E)$ flow field

To apply the boundary conditions given in equation (2.11) in a similar manner to equation (3.11), first one has to express $\left[\mathbf{u}^{(1)} \cdot \mathbf{e}_r\right]_s$, $\left[r\frac{\partial}{\partial r}\left(\mathbf{u}^{(1)} \cdot \mathbf{e}_r\right)\right]_s$ and $\left[\mathbf{r} \cdot \nabla \times \mathbf{u}^{(1)}\right]_s$ in terms of solid spherical harmonics in the following form (Happel & Brenner 1981)

$$\left[\mathbf{u}^{(1)} \cdot \mathbf{e}_r\right]_s = \sum_{n=-\infty}^{\infty}\left[\frac{n}{2N_\mu(2n+3)} p_n^{(1)} + n\Phi_n^{(1)}\right],$$
$$\left[r\frac{\partial}{\partial r}\left(\mathbf{u}^{(1)} \cdot \mathbf{e}_r\right)\right]_s = \sum_{n=-\infty}^{\infty}\left[\frac{n(n+1)}{2N_\mu(2n+3)} p_n^{(1)} + n(n-1)\Phi_n^{(1)}\right], \quad \text{(F1)}$$
$$\left[\mathbf{r} \cdot \nabla \times \mathbf{u}^{(1)}\right]_s = \sum_{n=-\infty}^{\infty} n(n+1) \chi_n^{(1)}.$$

To satisfy the shear stress balance, we have to obtain $\left[\mathbf{r} \cdot \nabla \times \left\{\mathbf{r} \times \left(\left(\boldsymbol{\tau}^{H(1)} + M\boldsymbol{\tau}^{E(1)}\right) \cdot \mathbf{e}_r\right)\right\}\right]_s$ and $\left[\mathbf{r} \cdot \nabla \times \left\{\left(\boldsymbol{\tau}^{H(1)} + M\boldsymbol{\tau}^{E(1)}\right) \cdot \mathbf{e}_r\right\}\right]_s$. Among these terms the hydrodynamic stress components can be obtained as

$$\left[\mathbf{r}\cdot\nabla\times\left(\mathbf{r}\times\left(\boldsymbol{\tau}^{H(1)}\cdot\mathbf{e}_r\right)\right)\right]_s = -N_\mu \sum_{n=-\infty}^{\infty}\left[2(n-1)n(n+1)\Phi_n^{(1)} + \frac{n^2(n+2)}{N_\mu(2n+3)}p_n^{(1)}\right], \quad \text{(F2)}$$

$$\left[\mathbf{r}\cdot\nabla\times\left(\boldsymbol{\tau}^{H(1)}\cdot\mathbf{e}_r\right)\right]_s = N_\mu \sum_{n=-\infty}^{\infty}(n-1)n(n+1)\chi_n^{(1)}. \quad \text{(F3)}$$

The radial traction vectors $\boldsymbol{\tau}_i^{E(1)}\cdot\mathbf{e}_r$ and $\boldsymbol{\tau}_e^{E(1)}\cdot\mathbf{e}_r$ arising at $O(Re_E)$ from the Maxwell stress tensor are expressed as

$$\boldsymbol{\tau}_i^{E(1)}\cdot\mathbf{e}_r = S\begin{bmatrix} E_{i,r}^{(0)}E_{i,r}^{(1)} - \frac{1}{2}\left(2E_{i,r}^{(0)}E_{i,r}^{(1)} + 2E_{i,\theta}^{(0)}E_{i,\theta}^{(1)} + 2E_{i,\phi}^{(0)}E_{i,\phi}^{(1)}\right) \\ \left(E_{i,r}^{(0)}E_{i,\theta}^{(1)} + E_{i,r}^{(1)}E_{i,\theta}^{(0)}\right) \\ \left(E_{i,r}^{(0)}E_{i,\phi}^{(1)} + E_{i,r}^{(1)}E_{i,\phi}^{(0)}\right) \end{bmatrix}, \quad \text{(F4)}$$

$$\boldsymbol{\tau}_e^{E(1)}\cdot\mathbf{e}_r = \begin{bmatrix} E_{e,r}^{(0)}E_{e,r}^{(1)} - \frac{1}{2}\left(2E_{e,r}^{(0)}E_{e,r}^{(1)} + 2E_{e,\theta}^{(0)}E_{e,\theta}^{(1)} + 2E_{e,\phi}^{(0)}E_{e,\phi}^{(1)}\right) \\ \left(E_{e,r}^{(0)}E_{e,\theta}^{(1)} + E_{e,r}^{(1)}E_{e,\theta}^{(0)}\right) \\ \left(E_{e,r}^{(0)}E_{e,\phi}^{(1)} + E_{e,r}^{(1)}E_{e,\phi}^{(0)}\right) \end{bmatrix}. \quad \text{(F5)}$$

We express the electrical stress components at the drop interface in terms of surface harmonics as

$$\begin{aligned}
\left[\mathbf{r}\cdot\nabla\times\left\{\mathbf{r}\times\left(\boldsymbol{\tau}_i^{E(1)}\cdot\mathbf{e}_r\right)\right\}\right]_s &= \sum_{n=0}^{\infty}\left(g_{n,m}^{i(1)}\cos m\phi + \hat{g}_{n,m}^{i(1)}\sin m\phi\right)P_{n,m}, \\
\left[\mathbf{r}\cdot\nabla\times\left\{\mathbf{r}\times\left(\boldsymbol{\tau}_e^{E(1)}\cdot\mathbf{e}_r\right)\right\}\right]_s &= \sum_{n=0}^{\infty}\left(g_{n,m}^{e(1)}\cos m\phi + \hat{g}_{n,m}^{e(1)}\sin m\phi\right)P_{n,m}, \\
\left[\mathbf{r}\cdot\nabla\times\left(\boldsymbol{\tau}_i^{E(1)}\cdot\mathbf{e}_r\right)\right]_s &= \sum_{n=0}^{\infty}\left(h_{n,m}^{e(1)}\cos m\phi + \hat{h}_{n,m}^{e(1)}\sin m\phi\right)P_{n,m}, \\
\left[\mathbf{r}\cdot\nabla\times\left(\boldsymbol{\tau}_e^{E(1)}\cdot\mathbf{e}_r\right)\right]_s &= \sum_{n=0}^{\infty}\left(h_{n,m}^{e(1)}\cos m\phi + \hat{h}_{n,m}^{e(1)}\sin m\phi\right)P_{n,m},
\end{aligned} \quad \text{(F6)}$$

where $g_{n,m}^{i(1)}, \hat{g}_{n,m}^{i(1)}, h_{n,m}^{i(1)}, \hat{h}_{n,m}^{i(1)}, g_{n,m}^{e(1)}, \hat{g}_{n,m}^{e(1)}, h_{n,m}^{e(1)}$ and $\hat{h}_{n,m}^{e(1)}$ are the coefficients that can be obtained after solving electrical potential distribution. The nonzero coefficients obtained after solving for the $O(Re_E)$ electric potential (equation (3.2)) are given in table 2. The complete expressions of the terms given in table 2 are extremely cumbersome to write down here. We provide here the expressions necessary for determining the drop velocity:

| $g_{n,m}^{i(1)}$ and $\hat{g}_{n,m}^{i(1)}$ | $g_{n,m}^{e(1)}$ and $\hat{g}_{n,m}^{e(1)}$ | $h_{n,m}^{i(1)}$ and $\hat{h}_{n,m}^{i(1)}$ | $h_{n,m}^{e(1)}$ and $\hat{h}_{n,m}^{e(1)}$ |
|---|---|---|---|
| $g_{1,0}^{i(1)}, g_{1,1}^{i(1)}, g_{2,0}^{i(1)}, g_{2,1}^{i(1)},$ | $g_{1,0}^{e(1)}, g_{1,1}^{e(1)}, g_{2,0}^{e(1)}, g_{2,1}^{e(1)},$ | $h_{2,0}^{i(1)}, h_{2,1}^{i(1)}, h_{2,2}^{i(1)}, h_{3,0}^{i(1)},$ | $h_{1,0}^{e(1)}, h_{1,1}^{e(1)}, h_{2,0}^{e(1)}, h_{2,1}^{e(1)},$ |
| $g_{2,2}^{i(1)}, g_{3,0}^{i(1)}, g_{3,1}^{i(1)}, g_{3,2}^{i(1)},$ | $g_{2,2}^{e(1)}, g_{3,0}^{e(1)}, g_{3,1}^{e(1)}, g_{3,2}^{e(1)},$ | $h_{3,1}^{i(1)}, h_{3,2}^{i(1)}, h_{3,3}^{i(1)}, \hat{h}_{2,1}^{i(1)},$ | $h_{2,2}^{e(1)}, h_{3,0}^{e(1)}, h_{3,1}^{e(1)}, h_{3,2}^{e(1)},$ |
| $g_{4,0}^{i(1)}, g_{4,1}^{i(1)}, g_{4,2}^{i(1)}, g_{4,3}^{i(1)},$ | $g_{4,0}^{e(1)}, g_{4,1}^{e(1)}, g_{4,2}^{e(1)}, g_{4,3}^{e(1)},$ | $\hat{h}_{2,2}^{i(1)}, \hat{h}_{3,1}^{i(1)}, \hat{h}_{3,2}^{i(1)}, \hat{h}_{3,3}^{i(1)}$ | $h_{3,3}^{e(1)}, \hat{h}_{1,1}^{e(1)}, \hat{h}_{2,1}^{e(1)}, \hat{h}_{2,2}^{e(1)},$ |
| $g_{4,4}^{i(1)}, \hat{g}_{1,1}^{i(1)}, \hat{g}_{2,1}^{i(1)}, \hat{g}_{2,2}^{i(1)},$ | $g_{4,4}^{e(1)}, \hat{g}_{1,1}^{e(1)}, \hat{g}_{2,1}^{e(1)}, \hat{g}_{2,2}^{e(1)},$ | | $\hat{h}_{3,1}^{e(1)}, \hat{h}_{3,2}^{e(1)}, \hat{h}_{3,3}^{e(1)}$ |
| $\hat{g}_{3,1}^{i(1)}, \hat{g}_{3,2}^{i(1)}, \hat{g}_{4,1}^{i(1)}, \hat{g}_{4,2}^{i(1)},$ | $\hat{g}_{3,1}^{e(1)}, \hat{g}_{3,2}^{e(1)}, \hat{g}_{4,1}^{e(1)}, \hat{g}_{4,2}^{e(1)},$ | | |
| $\hat{g}_{4,3}^{i(1)}, \hat{g}_{4,4}^{i(1)}$ | $\hat{g}_{4,3}^{e(1)}, \hat{g}_{4,4}^{e(1)}$ | | |

TABLE 2. Non-zero $g_{n,m}^{i(1)}, \hat{g}_{n,m}^{i(1)}, h_{n,m}^{i(1)}, \hat{h}_{n,m}^{i(1)}, g_{n,m}^{e(1)}, \hat{g}_{n,m}^{e(1)}, h_{n,m}^{e(1)}$ and $\hat{h}_{n,m}^{e(1)}$ obtained from the $O(Re_E)$ electric field.

$$g_{1,0}^{i(1)} = \frac{3S}{5(R+2)}\left(3E_x a_{2,1}^{(1)} + 3E_y \hat{a}_{2,1}^{(1)} + 2E_z a_{2,0}^{(1)}\right),$$

$$g_{1,1}^{i(1)} = -\frac{3S}{5(R+2)}\left(E_x a_{2,0}^{(1)} - 3E_z a_{2,1}^{(1)}\right),$$

$$\hat{g}_{1,1}^{i(1)} = -\frac{3S}{5(R+2)}\left(-3E_z \hat{a}_{2,1}^{(1)} + E_y a_{2,0}^{(1)}\right),$$

$$g_{1,0}^{e(1)} = \frac{9(R+1)}{5(R+2)}\left(3E_y \hat{b}_{-3,1}^{(1)} + 2b_{3,0}^{(1)} E_z + 3E_x b_{-3,1}^{(1)}\right), \quad \text{(F7)}$$

$$g_{1,1}^{e(1)} = \frac{9(R+1)}{5(R+2)}\left(3E_z b_{-3,1}^{(1)} - b_{-3,0}^{(1)} E_x\right),$$

$$\hat{g}_{1,1}^{e(1)} = \frac{9(R+1)}{5(R+2)}\left(3E_z \hat{b}_{-3,1}^{(1)} - E_y b_{3,0}^{(1)}\right).$$

Now, we obtain the coefficients of the solid spherical harmonics in terms of the unknown drop velocity at $O(Re_E)$ as:

$$A_{n,m}^{(1)} = \frac{(2n+3)}{n(2n+1)(\lambda+1)}\left[(2n+1)(2n-1)\beta_{n,m}^{(1)} + M\left(g_{n,m}^{i(1)} - g_{n,m}^{e(1)}\right)\right], \quad \text{(F8)}$$

$$B_{n,m}^{(1)} = -\frac{A_{n,m}^{(1)}}{2(2n+3)}, \quad \text{(F9)}$$

$$C_{n,m}^{(1)} = \frac{M\left(h_{n,m}^{e(1)} - h_{n,m}^{i(1)}\right)}{n(n+1)(\lambda(n-1)+(n+2))}, \tag{F10}$$

$$A_{-n-1,m}^{(1)} = -\frac{1}{(n+1)(2n+1)(\lambda+1)}\left[\left(4n^2-1\right)\{(2n+1)\lambda+2\}\beta_{n,m}^{(1)} + M(1-2n)\left(g_{n,m}^{i(1)} - g_{n,m}^{e(1)}\right)\right], \tag{F11}$$

$$B_{-n-1,m}^{(1)} = -\frac{(4n^2-1)\lambda\beta_{n,m}^{(1)} + M\left(g_{n,m}^{e(1)} - g_{n,m}^{i(1)}\right)}{2(2n+1)(n+1)(\lambda+1)}, \tag{F12}$$

$$C_{-n-1,m}^{(1)} = C_{n,m}^{(1)}, \tag{F13}$$

where $\beta_{1,0}^{(1)} = -U_{dz}^{(1)}, \beta_{1,1}^{(1)} = -U_{dx}^{(1)}$ and $\hat{\beta}_{1,1}^{(1)} = -U_{dy}^{(1)}$. Similarly, the coefficients $\hat{A}_{n,m}^{(1)}, \hat{B}_{n,m}^{(1)}, \hat{C}_{n,m}^{(1)}$, $\hat{A}_{-n-1,m}^{(1)}, \hat{B}_{-n-1,m}^{(1)}$ and $\hat{C}_{-n-1,m}^{(1)}$ can be obtained by replacing $\beta_{n,m}^{(1)}, g_{n,m}^{i(1)}$ and $g_{n,m}^{e(1)}$ by $\hat{\beta}_{n,m}^{(1)}, \hat{g}_{n,m}^{i(1)}$ and $\hat{g}_{n,m}^{e(1)}$ respectively in the above expressions.